\documentclass[runningheads,a4paper]{llncs}

\usepackage{amssymb}
\usepackage{graphicx}

\usepackage{url}
\urldef{\mailsa}\path|{alfred.hofmann, ursula.barth, ingrid.haas, frank.holzwarth,|
\urldef{\mailsb}\path|anna.kramer, leonie.kunz, christine.reiss, nicole.sator,|
\urldef{\mailsc}\path|erika.siebert-cole, peter.strasser, lncs}@springer.com|
\newcommand{\keywords}[1]{\par\addvspace\baselineskip
\noindent\keywordname\enspace\ignorespaces#1}

\begin{document}

\mainmatter  

\title{Challenges on Probabilistic Modeling for Evolving Networks}

\titlerunning{Challenges on Probabilistic Modeling for Evolving Networks}

%
%
%
%
%

\author{Jianguo Ding\inst{1}
\and Pascal Bouvry\inst{2}}

\authorrunning{Jianguo Ding, Pascal Bouvry}

\institute{Interdisciplinary Center for Security, Reliability and Trust\\ University of Luxembourg,
L-1359 Luxembourg\\
\email{Jianguo.Ding@ieee.org}\\
\and Faculty of Science, Technology and Communication\\University of Luxembourg,
L-1359 Luxembourg\\
\email{Pascal.Bouvry@uni.lu}
\\
}
\maketitle

\begin{abstract}

With the emerging of new networks, such as wireless sensor networks, vehicle networks, P2P networks, cloud computing, mobile Internet, or social networks, the network dynamics and complexity expands from system design, hardware, software, protocols, structures, integration, evolution, application, even to business goals. Thus the dynamics and uncertainty are unavoidable characteristics, which come from the regular network evolution and unexpected hardware defects, unavoidable software errors, incomplete management information and dependency relationship between the entities among the emerging complex networks. Due to the complexity of emerging networks, it is not always possible to build precise models in modeling and optimization (local and global) for networks. This paper presents a survey on probabilistic modeling for evolving networks and identifies the new challenges which emerge on the probabilistic models and optimization strategies in the potential application areas of network performance, network management and network security for evolving networks.

\end{abstract}

\section{Introduction}

It is recognized that three laws, Moore's law, Gilder's law and Metcalfe's law, which are governing the spread of technology and are related to the rapid evolution of IT networks \cite{[Din10]}. Moore's law indicates the computing capability of computers doubles every 18 months. Gilder's law claims the total bandwidth of communication systems triples every 12 months for the next 25 years. Metcalfe's law presents the value of a telecommunications networks is proportional to the square of the number of connected users of the systems $(n^{2})$.


The typical evolution in networks is paralleled with following changes: the improved/degraded hardware performance, updated software (system software, application software) and its functions, extended network structure with the integration of emerging heterogeneous networks (mobile communication networks, sensor networks, ad hoc networks, vehicle networks, overlay networks, and Internet of Things), extension of network scale with increasing wired/mobile and wireless devices joined, updated network protocols,  improved network functions (from information exchanging to complex online transaction and numerous services), dynamic functional evolution, variation of the network performance, and emerging network applications and network services.

The evolution of networks is not only on physical networks, but also on information networks and service networks, which is over physical networks. In real IT application scenarios, the evolution is characterized with the combinational evolving results on physical networks, information networks and service networks.

The evolving networks demonstrate the complex changes in network structure, network functions, network performance, and interoperation relationship with the time evolving, and thus more opportunistic networks and self-organizing networks come into being one trend of the network evolution. The emerging computing models (distributed computing, pervasive computing, cognitive computing, opportunistic computing, scalable computing, autonomic computing, physical computing, and probabilistic computing), which are employed to model and manage the complex dynamic networks and pertain to the operation, administration, maintenance, and provision of networked systems for secure (reliable) and effective network performance \cite{[Din12]}.

The dynamics and uncertainty are unavoidable characteristics, which come from the regular network evolution and unexpected hardware defects, unavoidable software errors, incomplete management information and dependency relationship between the entities among the emerging complex networks. Due to the complexity of emerging networks, it is not always possible to build precise models in modeling and optimization (local and global) for networks. New challenges emerge on the probabilistic models and optimization strategies in the areas of network performance, network management, network security for evolving IT networks.

This paper presents a systematic survey on the probabilistic modeling for evolving networks and identifies the new challenges which emerge on the probabilistic models and optimization strategies in the potential application areas of network performance, network management and network security for evolving networks.

\section{Emerging characteristics in evolving networks}

The evolving IT networks demonstrate emerging characteristics as follows:

\begin{enumerate}

  \item Dynamics

    \begin{itemize}

\item Some networks are running in dynamic style by nature, such as mobile communication networks, wireless sensor networks, vehicle networks, and overlay networks (P2P, VPN). These networks can be organized as opportunistic networks or self-organizing networks. That means the structure of the networks is changing over the topology, with the variation on routers, mobile servers and mobile clients.
\item The performance (robustness) of individual network components, such as routers, servers, clients or other key network devices/services vary with the network evolution. Some components are improved or degraded with the hardware performance, technical improvement, or systematic evolution.
\item The performance (robustness) of the individual links, which demonstrate the dependencies between main network components, changes with the structural or functional modification of the networks.
\item Both local and global changes are interdependent. That means any local changes may result in the variation of global network performance. Any global modification can result in the changes in local network performance as well.
\item Theoretically, the network function and structure have strong interdependence \cite{[New03]}. The evolving structure of networks will bring the changes in network functions. On the other side, it is possible that network function modification can result in the redesign/reconfiguration of network structure.

    \end{itemize}

  \item Heterogeneity
	
	In IT networks, there are 2 types of heterogeneous networks, integrated networks and overlay networks.

\begin{itemize}

 \item Integrated networks

With the advances of emerging networks, more heterogeneous networks are integrated. For example, the sensor networks are integrated with local networks, vehicle networks join mobile communication networks, heterogeneous network devices are integrated into Internet, and the trend of Internet of Things emerges. The heterogeneous network integration demonstrates the integration of different structures, different functions, different performance, different network protocols, different software components, and even different services. Integrated networks are not only the accumulation of networks, but also updated properties and functions merging with the evolution.

      \item Overlay networks

An overlay network is a computer network which is built on the top of another network. Nodes in the overlay networks can be connected by virtual or logical links, each of which corresponds to a path, perhaps through many physical links, in the underlying networks. Information networks and service networks based on the physical networks are commonly identified as overlay networks as well. Overlay networks organize peers with different strategies, thus their topology and routing performance are different. The consequent reliability and fault resiliency varies as well \cite{[LCP+05]}. Overlay networks are organized by spontaneous and dynamic connectivity between users/clients, this evolution model is accompanied with the continuing structure dynamics.

\end{itemize}

  \item Temporal Networks

Evolving is a time correlated process. The evolving network structure, describing how the network is wired and how the abstract nodes are connected, helps us to understand, predict and optimize the behaviour of dynamic networks. In many cases, however, the edges/links are not continuously active. In some cases, edges/links are active for non-negligible periods of time. Like network topology, the temporal structure of edge activations can affect dynamics of systems interaction through the network. The dynamic weights, which indicate the interdependencies between networks components, demonstrate the network evolving process as well. An evolving networks is a typical temporal network, which can be modelled to elucidate the behaviour of a dynamic system. The fundamental properties in temporal networks are  quite different from those for static networks.

  \item Complexity

Complexity has an important relationship to resilience and the robustness of systems, because resilience mechanisms such as self-organization and autonomic behaviour increase complexity, and increased complexity may result in greater network vulnerability \cite{[Str01]}. The complexity in evolving IT networks comes from structural complexity, network evolution, connection diversity, dynamical complexity, nodes diversity, meta-complication. Furthermore, the various complications can influence each other.

  \item Macro view vs. micro view

In evolving complex networks, self-organizing processes are deployed at multiple levels. Challenging questions about the dynamics of micro-macro transition include: (i) how are emergent properties related to micro interactions? (ii) how can we reverse-engineer the mechanics of complex system from their behaviour under a controlled set of external stimuli? Thus the interrelationship between micro and macro behaviour in evolving networks is rather important in application scenarios.

  \item Probability

In evolving network, the system dynamics and the intrinsic complexity make the complex networks with probabilistic properties. The incomplete and uncertain information need to be integrated into the research models, so that the system models can be more reasonable and realistic \cite{[Din08]}. Thus the time based probabilistic factors should be embedded into the network modeling in evolving networks.

\end{enumerate}

\section{Challenges for probabilistic modeling}

\begin{enumerate}

  \item Structure dynamics
	
In IT networks, the structure evolving is paralleled with the changes on:

\begin{itemize}
	\item Network nodes: New nodes (network hardware/software components) are added or removed from the network, or the improvement/degradation in the performance of nodes.
		\item Network links: New links (interoperation/interdependency) are added or removed from the network, or the improvement/degradation in the performance of links.
			\item Weight of the dependencies: The weight of the dependencies between network components indicates the measurement of the importance on the performance or dependencies among the related components. It is a time related function during the evolution.
\end{itemize}

Two types of dynamics are required in order to fully understand how networks evolve over time. The type of dynamics most commonly used in IT network analysis are those dealing with the nature of interactions between nodes as a consequence of the network structure, which is called \textbf{dynamics on the network}. This category governs how nodes react to each other based on the overall structure of the network. The second type of network dynamics, named \textbf{dynamics of the network}, governs the changes in the network structure and evolution of the structure.

There are three approaches to model the dynamic networks:

\begin{itemize}
	\item The evolution of a network can be described as a sequence of static networks and since there exist many parameters to describe accurately a static network, one can study the evolution of the network through the evolution of these parameters.
		\item  The evolution itself can be studied with defined parameters to capture the evolving properties, such as the rate of appearance or disappearance of nodes and edges.
			\item An intermediate approach can be used which consists specific phenomena in studying or users of interest with time.

\end{itemize}

The approach selection is based on the specific scenario. For example, if the network structure keeps stable and with minor modification, the sequence snapshot of the static networks can be considered as an appropriate evolving model. The network infrastructure and network backbone follow this class. But suppose the network structure evolves with great changes, and then the parameter modelling or intermediate approach might be appropriate. Mobile communication networks, sensor networks, vehicle networks belong to this class.


In modelling the evolving networks, not only the structure properties (topologies, nodes, and links) are included, but also the non-structural properties (weights, importance, functions) should be considered. It is also a challenge to model the dynamic structure and evolving properties in integrated heterogeneous network and overlay networks.

  \item Stochastic dynamics
	
Along the network evolution, some changes will generate new data characteristics which might not to be the property of the historical data. Generally, the building process of the network and the parameter estimation requires more data as the number of variables varies, as long as the accuracy in the estimations and in the network topology is to be maintained. However, the network evolution is not necessarily to maintain the accuracy in the parameter estimation because of the unstable and dynamic properties. There is some difficulties in obtaining a stable historical data for an evolving network on application scenarios.

  \item Model of heterogeneous networks
	
	Evolving networks are composed of heterogeneous networks in network structures and functions. This will make the integrated network model more complex, since some unbalanced and heterogenous network sections are not integrated consistently in following unique or common principles in the structure and the functions.

  \item Relationship between macro and micro networks
	
For a large scale heterogeneous network, the macro characteristics are strongly related to the micro pieces of the network and vice versa. When modeling the global probabilistic network, the local subnets modeling and their merging style between pieces are rather important, in which the interdependencies can be identified and measured reasonably. However, large scale complex networks demonstrate new properties which is hard to be identified from micro networks, such as small world property, scale free network, etc.

  \item Control and feedback
	
In modeling network application and services, the control and feedback loop is inevitable at different levels, particularly in logical and service networks. The loops among networks are apt to make the related modeling out of control. The overlap (repeat) dependencies vague the relationship between the networked components.

  \item Computing complexity
	
The complexity in emerging networks makes medium size models usually intractable, since the number of variables involved is greater than in static models. Highly connected networks and dynamic changes among the network structure and dependencies between related components make the evolving networks total complex dynamic systems, and thus brings very challenging problems in computing complexity.

\item Probabilistic factors

The dynamic and complex network behaviours inevitable brings probabilistic
factors to the evolving networks. Thus probabilistic factors should be included in the models of evolving networks. The combination of probabilistic
models and complex network model challenges the modelling of dynamic
evolving networks.

\end{enumerate}

\section{Probabilistic Modeling for Dynamic Networks}

The goal of modelling for evolving networks is to model the state of a system and
its evolution over time in a richer and more natural way. It is widely recognized
that probabilistic graphical models provide a good framework for both knowledge
representation and probabilistic inference for dynamic evolving networks.

A probabilistic dynamic model will be considered as a sequence of graphs indexed by the time, representing the temporal evolution of a system. Each graph
symbolizes the state of the system and the dependencies among its components
at a given time. The dynamic behaviour of the components of the system is described by a set of temporal dependencies among these components in different
time slices. Furthermore, these dependencies are quantified by conditional probability tables associated with the components of the system. In order to make the
management of such models feasible, a set of restrictions must be considered for
both its qualitative and quantitative aspects.

\subsection{Dynamic Graphs}

The dynamic behaviour of any specific system which changes over time requires
an implicit or explicit time representation. To model such systems is a very important task: the initial structure of the model and its propagation over time,
the probabilities attached to the structure, the qualitative and quantitative interrelations among variables in different time slices, etc., need to be taken into
account \cite{[LLL+96]}.

Basically, a network can be modelled as a graph, which includes essential elements: nodes, links, and weighs on links or/and nodes.
	
	A graph can be defined as a triple $(V,E,f_{V}, f_{E})$ where $V$ is
a set of vertexes, $E$ is a set of edges ${u,v}$, and $f$ is a function, $f_{V}: V\rightarrow N$, $f_{E}: E\rightarrow N$, where $N$ is some number system, assigning a value or a weight. Depending on the context, the weights may be real numbers, complex numbers, integers, elements of some group, etc.	
 A network or fully weighted graph has weights assigned to both nodes and edges.

These definitions of (static) graphs and networks involve the following entities: $V$ (a set of nodes), $E$ (a set of edges), $f_{V}$ (mapping vertexes to numbers), $f_{E}$ (mapping edges to numbers). A dynamic graph is obtained when any of these four entities changes over time. Thus, there are several basic kinds of dynamic graphs.

\begin{itemize}
\item in a node-dynamic graph or digraph, the set $V$ varies with time. Thus, some nodes may be added or removed. When nodes are removed, the edges incident with them are also eliminated.
	\item in an edge-dynamic (or arc-dynamic) graph or digraph, the set $E$ varies with time. Thus, edges may be added or deleted from the graph or digraph.
		\item in a node weighted dynamic graph, the function $f_{V}$ varies with time; thus, the weights on the nodes also change.
			\item in an edge weighted dynamic graph or digraph, the function $f_{E}$ varies with time.
				\item in fully weighted dynamic graph, both functions $f_{V}$ and $f_{E}$ may vary with time.
\end{itemize}

Thus a dynamic graph is defined as a triple with time parameter $t$: $(V^t,E^t,f_{V}^t, f_{E}^t)$. Harary classifies the dynamic graphs by the
change of any of these \cite{[HG97]}:

\begin{enumerate}
	\item Node dynamic graphs where the vertex set $V^t$
changes over time $t$.
	\item Edge dynamic graphs where the edge set $E^t$
change over time $t$.
	\item Node weighted dynamic graphs where the $f_{V}^t$ function
varies with time $t$.
	\item Edge weighted dynamic graphs where the $f_{E}^t$
function varies with time $t$.
\end{enumerate}

All combinations of the above types can occur. For example, a computer network with changing bandwidth (edge-weight), changing topology (edges being added or deleted), changing computing power (node-weights changing), and computers representing nodes crashing, and recovering represents a dynamic graph that entails all the above basic types.

Work on dynamic graph theory have been motivated by
finding patterns and laws. Power laws, small diameters,
shrinking diameters have been observed. Graph generation
models that try to capture these properties are proposed
to synthetically generate such networks \cite{[Bil08]}. There are several
problems to be answered in these complex dynamic networks.

\begin{itemize}
	\item Is the
network evolving normally?
		\item What is normal behaviour of the network?
			\item Is
there a phase transition along the network evolving?
\end{itemize}



There is a strong correlation between finding patterns in
static graphs and dynamic evolving graphs.
%
%

Graph similarity functions, which is used to measure the degree of the dynamics on networks,
are categorized into two groups:

\begin{itemize}
	\item feature based similarity measures
		\item structure based similarity measures
\end{itemize}

Using the topology of the graphs, two similarity metrics
have been defined, maximum common subgraph distance and
the graph edit distance.


Graph clustering has become a central tool for the analysis of dynamic networks in general, with applications
ranging from the field of social sciences to biology and to the growing field of complex systems. The
general aim of graph clustering is to identify dense subgraphs in networks. Countless formalizations
thereof exist, however, the overwhelming majority of algorithms for graph clustering relies on
heuristics, e.g., for some NP-hard optimization problem, and do not allow for any structural
guarantee on their output.

\subsection{Power Law Random Graphs}
	
Random graphs can date back to the work of Erd\H{o}s and R\'{e}nyi
for the theory of random
graphs \cite{[ER60]}. The random graph model $G(n, e)$ assigns
uniform probability to all graphs with $n$ nodes and $e$ edges while in the
random graph model $\mathcal{G}(n, p)$ each edge is chosen with probability $p$.

Power law random graph model \cite{[ACL01]} is an extension of random graph, whose degree distribution follows a power law. Most of IT network system have this properties. Power law rand graph model has two parameters. The two
parameters only roughly delineate the size and density but they are natural
and convenient for describing a power law degree sequence. The power law
random graph model $P(\alpha, \beta)$ is described as follows. Let $y$ be the number of
nodes with degree $x$. $P(\alpha, \beta)$ assigns uniform probability to all graphs with
$y = e^\alpha/x^\beta$ (where self loops are allowed). Note that $\alpha$ is the intercept and
$\beta$ is the (negative) slope when the degree sequence is plotted on a log-log
scale.

There is also an alternative power law random graph
model analogous to the uniform graph model $\mathcal{G}(n, p)$. Instead of having a fixed degree sequence, the random graph has an expected degree sequence
distribution. The two models are basically asymptotically equivalent, subject to bounding error estimates of the variances.

The power law random graph model provides an approach to model the dynamic evolving complex networks. However, there are some questions that remain to be resolved. For example,
what is the effect of time scaling? How does it correspond with the evolution
of $\beta$? What are the structural behaviours of the power law random graphs?

\subsection{Dynamic Flowgraph Methodology}

The dynamic flowgraph methodology (DFM) \cite{[BK11]} is an approach to model and analyze the
behaviour of dynamic systems for reliability/safety assessment and verification. DFM models
express the logic of the system in terms of causal relationships between physical variables and
states of the system. The time aspects of the system (execution of control commands, dynamics of
the process) are represented as a series of discrete state transitions. DFM can be used for
identifying how certain postulated events may occur in a system. The result is a set of timed fault
trees, whose prime implicants (multi-state analogue of minimal cut sets) can be used to identify
system faults resulting from unanticipated combinations of software logic errors, hardware
failures, human errors and adverse environmental conditions.

DFM models are directed graphs, analyzed by discrete time instances. They consist of
variable and condition nodes; causality and condition edges; and transfer and transition boxes
and their associated decision tables. A node represents a variable that can be in one of a finite
number of predefined states. The state of a node can change at discrete time instances. The state
of the node is determined by the states of its input nodes. Each node can have several inputs but
only one output (its state). The state of the node can act as an input to possibly several other
nodes. The state of a node at time $t$ is determined by

\begin{itemize}
	\item the states of its input nodes at a single instance of time (say, $t-n$),
	\item the lag $n$, an integer that tells how many time instances it takes for an input to cause
the state of the present node.
\end{itemize}

The state of a node, as a function of the states of its input nodes, is determined by a
decision table. A decision table is an extension of the truth table where each variable can be
represented with any finite number of states. The decision table contains a row for each possible
combination of input variable states. The maximum possible number of rows in the decision
table is the product of the numbers of states of the input nodes.

After construction, the DFM model can be analyzed in two different modes, deductive and
inductive. In inductive analysis, event sequences are traced from causes to effects; this
corresponds to simulation of the model. In deductive analysis, event sequences are traced
backward from effects to causes.

A deductive analysis starts with the identification of a particular system condition of interest
(a top event); usually this condition corresponds to a failure. To find the root causes of the top
event, the model is backtracked for a predefined number of steps through the network of nodes,
edges, and boxes. This means that the model is worked backward in the cause-and-effect flow to
find what states of variables (and at what time instances) are needed to produce the top event.
The result of a deductive analysis is a set of prime implicants.

A prime implicant consists of a set of triplets $(V, S, T)$; each triplet tells that variable $V$ is in
a state $S$ at time $T$. The circumstances described by the set of triplets cause the top event. Prime
implicants are similar to minimal cut sets of fault tree analysis, except that prime implicants are
timed and they deal with multi-valued variables (fault trees deal with Boolean variables). A
useful analogy is that deductive analysis corresponds to minimal cut set search of a fault tree.
Once primary implicants have been found, the top event probability can be quantified in a fault tree.

For large scale dynamic networks, the state analysis and the fault propagation among dynamic networks make the resolution intangible with huge computing complexity.

\subsection{Dynamic Factor Graphs}

Directed and undirected networks coexist in most IT networks.  For large networks (graphs), the factorization properties of a graphical model,
whether undirected or directed, may be difficult to visualize from the
usual depictions of graphs. The formalism of factor graphs provides an
alternative graphical representation, one which emphasizes the factorization
of the distribution.

Let $F$ represent an index set for the set of factors defining a graphical
model distribution. In the undirected case, this set indexes the
collection $C$ of cliques, while in the directed case $F$ indexes the set
of parent-child neighborhoods. We then consider a bipartite graph
$G = (V,E,F)$, where $V$ is the original set of vertexes, and $E$ is a new
edge set, joining only vertexes $s \in V$ to factors $a \in F$. In particular,
edge $(s,a) \in E$ if and only if $x_s$ participates in the factor indexed by
$a \in F$.

For undirected models, the factor graph representation is of particular value when $C$ consists of more than the maximal cliques. Indeed, the compatibility functions for the nonmaximal cliques do not have an explicit representation in the usual representation of an undirected graph. However, the factor graph makes them explicit.

Time series collected from real-world phenomena are often an incomplete picture
of a complex underlying dynamical process with a high-dimensional state that
cannot be directly observed.

%

The simplest approach to modeling time series relies on time-delay embedding:
the model learns to predict one sample from a number of past samples with
a limited temporal period. This method can use linear auto-regressive models, as
well as non-linear ones based on kernel methods (e.g. support-vector regression), neural networks (including convolutional networks such as time
delay neural networks), and other non-linear regression models. Unfortunately,
these approaches have difficult in capturing hidden dynamics with
long-term dependency because the state information is only accessible indirectly
through a (possibly very long) sequence of observations \cite{[BSF94]}.

To capture long-term dynamical dependencies, the model must have an internal
state with dynamical constraints that predict the state at a given time
from the states and observations at previous times (e.g. a state-space model).
In general, the dependencies between state and observation variables can be expressed
in the form of a Factor Graph for sequential data, in which a graph
motif is replicated at every time step.

For a complex and non-linear system, a model might allow the use of complex
functions to predict the state and observations, and will sacrifice the probabilistic
nature of the inference. Instead, the inference process (including during learning)
will produce the most likely (minimum energy) sequence of states given the
observations. Dynamic Factor Graph (DFG) is a natural
extension of Factor Graphs specifically tuned for sequential data.
To model complex dynamics, DFG allows the state at a given
time to depend on the states and observations over several past time steps.

Dynamical Factor Graphs manage to perfectly reconstruct multiple
oscillatory sources or a multivariate chaotic attractor from an observed one-dimensional
time series. DFGs also outperform Kalman Smoothers and other
neural network techniques on a chaotic time series prediction tasks, DFGs can be used for the estimation of missing
motion capture data. Proper regularization such as smoothness or a sparsity
penalty on the parameters enable to avoid trivial solutions for high-dimensional
latent variables \cite{[ML09]}.

\subsection{Time-varying Graphs}

Time-varying graphs (TVG) have been a topic of active research
recently in the study of communication
networks with intermittent connectivity such as delay-tolerant
networks and even disruption-tolerant social networks; duty cycling wireless sensor networks,
and so on. Existing research on time-varying graphs ranges
from algorithmic studies on graph journeys to analysis of
specific properties such as flooding time in dynamic random
graphs. Empirical simulation-based analysis of certain
temporal graph properties such as temporal distance and temporal
efficiency are hot topics in this area.

%

The TVG can describe a multitude of different scenarios, from transportation
networks to communication networks, complex systems, or social networks. Some research questions are generated by the application requirements in dynamic networks.

One important task is to explore the universe of dynamic networks using the
formal tools provided by the TVG formalism. The long-term goal is to provide a comprehensive
map of this universe, to identify both the commonality and the natural differences between the various
types of dynamical systems modeled by TVG \cite{[CFQ+11]}.

The design and analysis of distributed algorithms
and protocols for time-varying graphs is an open research area. In fact very few problems have been
attacked so far: routing and broadcasting in delay-tolerant networks; broadcasting and exploration in
opportunistic-mobility networks; new self-stabilization techniques; detection of
emergence and resilience of communities, and viral marketing in social networks.

If the interactions in a network can be planned and decided by a
designer, then a number of new interesting optimization problems arise with the design of time-varying
graph. They may concern, for example, the minimization of the temporal diameter or the balancing
of nodes eccentricities.

Analyzing the complexity of a distributed algorithm in a TVG , e.g. in number
of messages, is not trivial, partly because contrarily to the static cases, the complexity of an algorithm
in a dynamic network has a strong dependency, not only on the usual network parameters (number of
nodes, edges, etc.), but also on the number of topological events taking place during its execution. In
many of the algorithms, the majority of messages is in fact directly triggered by
topological events, e.g., in reaction to the local appearance or disappearance of an edge. The number of
topological events therefore represents a new complexity parameter, whose impact on various problems
remains to study.


Through
the use of the interaction-centric point view, TVGs enable to look at the interplay between topological
aspects that allow local interaction to have global effects.

%
%
%
%
%

\subsection{Dynamic Bayesian Networks}

Dynamic Beyesian Networks (DBN), which is an extension of causal probabilistic networks \cite{[Ber89]} and static Bayesian networks, is to model a system that is dynamically
changing or evolving over time. This model will enable users to monitor and
update the system as time proceeds, and even predict further behaviour of the
system. In every time slice of a temporal model corresponds to one particular
state of a system, and if the movement between the slices reflects a change in
state instead of time.

Dynamic Beyesian Networks are usually defined as special case of singly
connected Bayesian networks specifically aimed at time series modelling. 

All
the nodes, edges and probabilities that form static interpretation of a system is
identical to a Bayesian network. Variables can be denoted as the sate of a DBN,
because they include a temporal dimension. The states of any system described
as a DBN satisfy the Markovian condition, that is defined as follows: The sate
of a system at time $t$ depends only on its immediate past, i.e. its state at time $t-1$. Also, this property is frequently considered as a definition of First order
Markov property: the future is independent of the past given the present.
The states of a dynamic model do not need to be directly observable. They
may influence some other variables that we can directly measure or calculate.
Also, the state of some system needs not to be a unique, simple state. It may be
regarded as a complex structure of interacting states. Each state in a dynamic
model at one time instance may depend on one or more states at the previous
time instance or/and on some states in the same time instance \cite{[Mur02]}. So, in DBN
states of a system at time $t$ may depend on system states at time $t-1$ and possibly on current states of some other nodes in the fragment of DBN structure that represents variables at time $t$.

It is not easy to model time and uncertainty in a way that clearly and adequately represents the problem domains at hand. Related approaches can be classified into three broad categories:

\begin{itemize}
	\item Models that use static BNs and formal grammars to represent temporal dimension (known as Probabilistic Temporal Networks)
		\item Models that use mixture of probabilistic and non-probabilistic frameworks
			\item Models that introduce temporal nodes into static BNs structure to represent time dependence.
\end{itemize}

%

By using a DBN,
we assume that dynamic data are generated sequentially by some hidden states of
a dynamic factor evolving over time. Since the hidden states cannot be observed directly,
they can only be inferred from the observed data given a learned DBN. Learning
a DBN involves estimating both its structure and parameters from data \cite{[TL03]}. The structure of
a DBN refers primarily to (1) the number of hidden states of each hidden variable in a
model and (2) the conditional dependence among hidden states of all the hidden variables
of a model, i.e, factorization of the model state space for determining the topology of a
graph network. There have been extensive studies in the machine learning community on
efficient parameter learning when the structure of a model is known a priori. Mixed atemporal and temporal independence relations among DBN models is examined as well \cite{[IL07]}. However, much less efforts have been made to tackle the more challenging problem
of learning the optimal structure of an unknown DBN. As a consequence,
most previous DBN-based data modelling approaches avoid the structure learning
problem by setting the structure manually. However, it has been shown that a
learned structure can be advantageous over those that are manually set \cite{[GX03]}.

DBNs represent the state of the world as a set of variables, and
model the probabilistic dependencies of the variables within
and between time steps. While a major advance over previous
approaches, DBNs are still unable to compactly represent
many real-world domains. In particular, domains can contain
multiple objects and classes of objects, as well as multiple
kinds of relations among them; and objects and relations can
appear and disappear over time. Capturing such a domain
in a DBN would require exhaustively representing all
possible objects and relations among them. This raises two
problems. The first one is that the computational cost of using
such a DBN would likely be prohibitive. The second one is
that reducing the rich structure of the domain to a very large
``flat" DBN would render it essentially incomprehensible to human beings \cite{[MP01]}.


\subsection{Probabilistic Complex Networks}

The term of ``complex network" is usually used for referring the natural networks that are usually complex and cannot be modeled just through random
graphs. Most real-world networks have these complex topological features, such
as, heavy-tail in the degree distribution, high clustering coefficient, assortativity
or disassortativity among vertexes, inherent multiparty structure, self-similar
hierarchical structure, etc. Clustering coefficient represents the ratio of a network that satisfies your friends are also mutually friends. The assortativity
represents the grouping of nodes. Inherent multiparty structure indicates that
there exists many intrinsic multiparty properties in the real world.

On the contrary, simple networks usually have these properties. For instance,
they can be represented by graphs such as a lattice or random graph. The topological structure is roughly the same in any part of network. And, they do not
posses the above complex network features. Examples of complex networks include social networks, computer networks, biological networks - neurons, or protein structure, river networks, power-line networks, etc.

A probabilistic complex network can be defined as a set of probabilistic nodes
and probabilistic edges in a network topology which follows the characteristics
of complex networks, such as power-law of the degree distribution of nodes and
the small world phenomenon which specifies the shortest path between any two
nodes are generally small. The network topology can be represented by setting
the probabilities to zero in some of the edges at a completely connected network.
A probabilistic node means there are possibly either discrete states or continuous
value (or attribute vector) in the node. Similarly, a probabilistic edge represents
that the discrete state or value of edges is probabilistic \cite{[Lin07]}.

In a probabilistic complex network, there should be causal
relationships between the states/values of nodes and edges.
This is the biggest difference between a probabilistic complex
network and a random graph. The inter-nodes, inter-edges, and
nodes-edges state relationships can be determined by a
deterministic or probabilistic model by some pre-specified rules.
When defining dynamic probabilistic complex network
as a probabilistic complex network, evolves over time will be considered. The
probably distributions of nodes and edges can be different at any
sampled time. For simplicity, in IT networks, we only consider the discrete cases
for probabilistic complex networks.

Many real-life network systems (such as Internet, WWW, etc.) can be modeled as probabilistic complex networks of interacting components. Although the study of such large scale
networks is not new, there has recently been much renewed
interest in this field. This is due to technological advances of two
types: (i) the collection of data which depict large networks in
 detail, and (ii) the development of computational
tools for the analysis of data. Among the well-studied examples
of such networks are the World Wide Web, citation networks,
neuronal connections, metabolic networks, ecological webs and more \cite{[AB02]} .

Traditional Erd\H{o}s and R\'{e}nyi random graph models have possion
degree distributions. However, it has been found that
many real life networks follow power law distributions. Generalized random graph models have been
proposed to mimic the power law degree distribution of the
real networks but these models do not explain how such
a phenomena occurs in these graphs. Barabasi et al.\cite{[Bar99]}
introduced a model (BA Model) the concept of preferential attachment for this
purpose.

The BA model is an algorithm for generating random scale-free networks using a preferential attachment mechanism. Scale-free networks are widely observed in natural and human-made systems, including the Internet, the world wide web, citation networks, and some social networks.

The network begins with an initial network of $m_0$ nodes ($m_0 \geq 2$) and the degree of each node in the initial network should be at least 1, otherwise it will always remain disconnected from the rest of the network.
New nodes are added to the network one at a time. Each new node is connected to  existing nodes with a probability that is proportional to the number of links that the existing nodes already have. Formally, the probability $p_i$ that the new node is connected to node $i$ is

\begin{equation}
   p_i=\frac{k_i}{\Sigma_jk_j}
\end{equation}

where $k_i$ is the degree of node i and the sum is made over all preexisting nodes $j$. Heavily linked nodes (``hubs") tend to quickly accumulate even more links, while nodes with only a few links are unlikely to be chosen as the destination for a new link. The new nodes have a ``preference" to attach themselves to the already heavily linked nodes.

Follow the BA model, there are two types of sub-models:

\begin{itemize}
	\item Model A retains growth but does not include preferential attachment. The probability of a new node connecting to any pre-existing node is equal. The resulting degree distribution in the limit is geometric \cite{[PRR13]}, indicating that growth alone is not sufficient to produce a scale-free structure.
	\item Model B retains preferential attachment but eliminates growth. The model begins with a fixed number of disconnected nodes and adds links, preferentially choosing high degree nodes as link destinations. Though the degree distribution early in the simulation looks scale-free, the distribution is not stable, and it eventually becomes nearly Gaussian as the network nears saturation. So preferential attachment alone is not sufficient to produce a scale-free structure.
The failure of models A and B to lead to a scale-free distribution indicates that growth and preferential attachment are needed simultaneously to reproduce the stationary power-law distribution observed in real networks \cite{[AB02]}.
\end{itemize}

In modelling probabilistic complex networks, still some challenges forward:

\begin{enumerate}
	\item  Finding dynamic community structures in a probabilistic complex networks,

	\item  Investigating appropriate classification techniques,

	\item  Applying on various practical datasets,

    \item  Processing large datasets,

	\item  Investigating the probabilistic approach to do information classification and network topology learning.

\end{enumerate}

\section{Conclusions}

With the evolution in network structures and the applications, more requirements are generated for the modelling of
probabilistic dynamic networks. The challenges include structure dynamics, probabilistic factors, heterogeneous networks,
control and feedback, computing complexity, etc. There are several approaches in modelling the dynamic and probabilistic
factors and holistic evolving networks, such as  dynamic graph, power law random graph, dynamic flowgraph method, dynamic
factor graphs modeling, time-varying graphs, dynamic Bayesian Networks, probabilistic complex network modelling. However,
most of the approaches are linked to application scenarios and focus on specific dynamic systems and on particular dynamic
behaviours. There is no common approach available to deal with the various dynamics among the evolving networks. 

Based on the observation and requirements on empirical research, the following topics are deserved for detailed investigation:
  
  \begin{itemize}

    \item dynamic properties of the network should be identified. This includes the network structure changes, the scale, degree and speed of the changes among networks;
    \item the direct/indirect (hidden) factors, which contribute to the network evolving, should be traced and identified;
    \item the trend/principle of the evolving should be identified with appropriate approaches and try to make the future status of the evolving networks be predictable. This mainly depends on specific application scenarios (datasets);
    \item local (micro) and global (macro) changes on evolving networks should be distinguished and synthesized, so that the dynamics of the evolving networks can be examined systematically;
    \item control and optimization on dynamic networks is important issues to adjust the network performance;
    \item  the computing complexity should be controllable and feasible, particularly in dealing with large scale and probabilistic data.
 \end{itemize}
  
  Thus more and efficient approaches and strategies should be developed to resolve the challenging problems and to improve the dynamic modelling of modern evolving networks.

\end{document}